\documentclass[aps,prl,twocolumn,superscriptaddress,showpacs]{revtex4-1}
\usepackage{amssymb}
\usepackage{amsmath}
\usepackage{bm}
\usepackage{url}
\usepackage{graphicx}
\DeclareMathOperator{\Tr}{Tr}

\begin{document}
\title{Alignment of non-spherical active particles in chaotic flows}

\author{M. Borgnino}
\affiliation{Dipartimento di Fisica and INFN, Universit\`a di Torino, 
via P. Giuria 1, 10125 Torino, Italy}

\author{K. Gustavsson}
\affiliation{Department of Physics, Gothenburg University, 41296 Gothenburg, Sweden}

\author{F. De Lillo}
\affiliation{Dipartimento di Fisica and INFN, Universit\`a di Torino, 
via P. Giuria 1, 10125 Torino, Italy}

\author{G. Boffetta}
\affiliation{Dipartimento di Fisica and INFN, Universit\`a di Torino, 
via P. Giuria 1, 10125 Torino, Italy}

\author{M. Cencini}
\thanks{Corresponding author}
\email{massimo.cencini@cnr.it}
\affiliation{Istituto dei Sistemi Complessi, CNR, via dei Taurini 19, 00185 Rome, Italy
and  INFN, sez. Roma2 ``Tor  Vergata''}

\author{B. Mehlig} \affiliation{Department of Physics, Gothenburg
  University, 41296 Gothenburg, Sweden}

\begin{abstract}
We study the orientation statistics of spheroidal, axisymmetric
microswimmers, with shapes ranging from disks to rods, swimming in
chaotic, moderately turbulent flows.  Numerical
  simulations show that rod-like active particles preferentially align
  with the flow velocity.  To explain the underlying mechanism we
solve a statistical model via perturbation theory. We show that such
alignment is caused by correlations of fluid velocity and its
gradients along particle paths combined with fore-aft
  symmetry breaking due to both swimming and particle
nonsphericity. Remarkably, the discovered alignment
  is found to be a robust kinematical effect, independent of the
  underlying flow evolution. We discuss its possible relevance for
  aquatic ecology.
\end{abstract}


\maketitle 

Active particles, such as motile microorganisms or artificial
microswimmers, swim in a surrounding flow, either externally imposed
or self-generated.  In addition to transporting the active particles,
velocity gradients change their swimming direction by exerting a
shape-dependent torque \cite{jeffery1922,pedley1987}.  The complex
interplay of flow advection, particle orientation and self-propulsion
is fundamental to understand key processes in aquatic ecology
\cite{luchsinger1999,kiorboe2008,rusconi2015,taylor2012,guasto2012},
active matter modeling \cite{marchetti2013,bechinger2016,elgeti2015},
and nano/micro-technology with application to drug delivery
\cite{dreyfus2005,nelson2010}.

Even simple laminar steady flows give rise to intriguing phenomena
when combined with self-propulsion
\cite{torney2007transport,zottl2012,rusconi2014,junot2019swimming}.
Rod-shaped motile bacteria are expelled by vortices
\cite{torney2007transport} and display complex trajectories in pipe
flows \cite{zottl2012}.  Microfluidic experiments in shear flows found
that bacteria tumble in high shear regions, causing accumulation and
chemotactic depletion \cite{rusconi2014}. In shear flows, a different
tumbling mechanism traps bottom-heavy gyrotactic phytoplankton
\cite{durham2009}.  It has recently been found that individual
bacteria in steady porous flow can orient their swimming direction
with the local velocity leading to a strong enhancement (depletion) of
the dispersion along (transverse to) the mean flow direction
\cite{dehkharghani2019bacterial}.

The behavior of active particles in unsteady flows is considerably
less explored.  Gyrotactic swimmers form small-scale fractal patches
in turbulence
\cite{durham2013,zhan2014accumulation,gustavsson2016gyrotaxis},
sampling different flow regions depending on their shape
\cite{gustavsson2016gyrotaxis,borgnino2018shape}. Elongated swimmers,
such as bacteria, remain quite homogeneously distributed in turbulent
flows, while their orientation tends to nematically align with the
vorticity \cite{zhan2014accumulation,pujara2018rotations}, similarly
to elongated tracers
\cite{voth2017anisotropic,pumir2011orientation}. Much less is known
about their orientation with respect to the flow velocity, which is
key to light scattering in aquatic environments
\cite{boss,seymour2011microbizal}, and for the encounter rates between
organisms \cite{kiorboe2008}. For instance, flow reorientation of
elongated prey in the feeding currents of predators can strongly
modify the capture rates \cite{visser}. Moreover, flow induced changes
in the swimming direction can strongly alter chemotaxis, as found in
steady shear flows \cite{rusconi2014}.

\begin{figure*}[t!]
\centering
\includegraphics[width=1\textwidth]{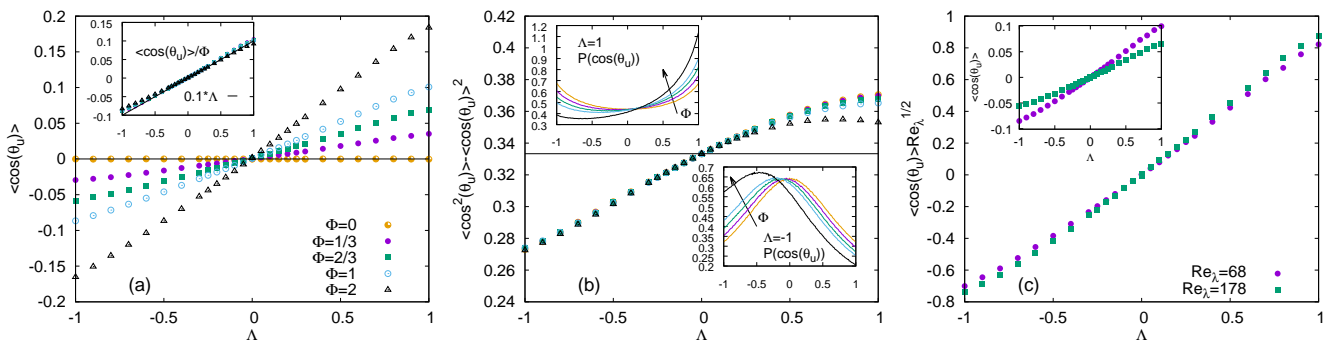}
\vspace{-0.5truecm}
\caption{(Color online) Statistics of particle orientation with
  respect to the flow velocity, obtained from DNS of the NSE (\ref{eq:ns}),
  as a function of the particle shape parameter $\Lambda$, for
  different swimming number $\Phi$.  (a) $\langle \cos\theta_u\rangle$
  vs $\Lambda$ for different $\Phi$ at $Re_\lambda\approx 68$.  Inset:
  $\langle \cos\theta_u\rangle/\Phi$ vs $\Lambda$, the solid line
  represents a linear best fit.  (b) Variance of $\cos\theta_u$ vs
  $\Lambda$ for different $\Phi$.  Top (bottom) inset shows the PDF of
  $\cos\theta_u$ for rod- (disk-) like particles with $\Phi$ from $0$ to
  $2$ along the arrows.  (c) Same as panel (a) for $\Phi=1$ and
  $Re_\lambda=68$ and $178$. Main panel shows $\langle
  \cos\theta_u\rangle \mathrm{Re}_\lambda^{1/2}$ vs $\Lambda$. Inset
  shows the non-rescaled data.  Data are obtained by averaging over
  100 snapshots, separated by about half large-scale eddy turnover
  time, with up to $3\cdot 10^5$ particles for each $\Lambda$ and
  $\Phi$ values.
\label{fig1}}
\end{figure*}

In this Letter, aiming to fill this gap, we investigate the dependence
of the orientation statistics of active particles on their shape and
speed in unsteady, moderately turbulent and stochastic flows.  We find
that swimming directions preferentially align with or against the
local velocity field depending on the particle shape. Solving, by
means of perturbative methods, the problem with a stochastic velocity
field we trace back the origin of such an alignment to the correlation
between flow velocity and its gradients along the particle path.

We consider dilute suspensions, disregarding any form of particle
interaction. In this limit, we can neglect flow modifications induced
by the active particles.  We model a microswimmer as a small,
neutrally buoyant, non-spherical, axisymmetric particle swimming with
constant speed, $v_\mathrm{s}$, in the direction $\bm n$ of its
symmetry axis.  Assuming the particle size is smaller than the
smallest flow scale, the particle center of mass $\bm x$ evolves as
\cite{pedley1987}
\begin{equation}
\dot{\bm x} =
\bm u(\bm x,t) + v_\mathrm{s} \bm n\,, 
\label{eq:dotx}
\end{equation}
 $\bm u(\bm x,t)$ being the fluid velocity at the particle position.
Particle orientation  rotates in response to velocity gradients, according to  Jeffery's dynamics \cite{jeffery1922}
\begin{equation}
\dot{\bm n} = [\mathbb{O}(\bm x,t) + \Lambda \mathbb{S}(\bm x,t)]\bm n 
-\Lambda [\bm n\cdot\mathbb{S}(\bm x,t)\bm n]\bm n \equiv \bm J(\bm n) \,,
\label{eq:dotn}
\end{equation}
where $\mathbb{O}$ and $\mathbb{S}$ are the antisymmetric (vorticity)
and symmetric (strain) components of the velocity gradient matrix
$\mathbb{A}_{ij}=\partial u_i/\partial x_j$, respectively.
For $v_s\!=\!0$, the above dynamics reproduces that
  of spheroidal tracers that have recently gathered much attention
  \cite{voth2017anisotropic,pumir2011orientation,gustavsson2015particle}.

The dynamics is controlled by two dimensionless numbers. The first is
the \textit{shape parameter} $\Lambda=(a^2-b^2)/(a^2+b^2)$ ($a$ and
$b$ being the particle size along and perpendicularly to the symmetry
axis): $\Lambda\!=\!0$ for spheres, and $\Lambda\!=\!\pm 1$ for
infinitely slender rods and thin platelets, respectively. The second
is the \textit{swimming number} $\Phi=v_{\rm s}\tau/\ell$, $\ell$ and
$\tau$ being the flow typical scale and time, discussed below.  For
$\Lambda>0$, Eqs.~(\ref{eq:dotx}-\ref{eq:dotn}) represent a minimal
model for a smooth swimming (not tumbling) bacterium
\cite{rusconi2014,junot2019swimming}.  Rotational diffusivity in
Eq.~(\ref{eq:dotn}) is neglected to reduce the number of parameters.

We start considering homogeneous, isotropic turbulent
flows obtained by direct numerical simulations (DNS) of the
Navier-Stokes equations (NSE)
\begin{equation}
\partial_t \bm u+\bm u\cdot \bm \nabla \bm u= -\bm \nabla p +\nu \Delta \bm u +\bm F\,, \label{eq:ns}
\end{equation}
where fluid density is scaled to unity, pressure $p$ ensures flow
incompressibility ($\bm \nabla\cdot\bm u\!=\!0$) and $\nu$ is the
viscosity. The stirring force $\bm F$ is an incompressible, zero-mean,
temporally uncorrelated Gaussian random field, injecting kinetic
energy at large scales at a rate $\epsilon$ to generate a
statistically steady state. We solve Eqs.~(\ref{eq:ns}), by means of a
$2/3$-dealiased pseudospectral solver with a $2^{nd}$ order
Runge-Kutta scheme, in a triply periodic domain with
$N^{3}\!=\!128^3\!-\!512^{3}$ mesh points.  The Kolmogorov length,
$\eta=(\nu^3/\epsilon)^{1/4}$, is larger than the grid spacing and the
time step much smaller than the Kolmogorov timescale,
$\tau_\eta=(\nu/\epsilon )^{1/2}$, to well resolve the small scales
dynamics.  Velocity and its gradients at particle positions, needed to
integrate Eqs.~(\ref{eq:dotx}-\ref{eq:dotn}), are obtained via a third
order interpolation scheme.  The swimming number is defined as
$\Phi=v_s \tau_\eta/\eta=v_s/u_\eta$, $u_\eta$ being the Kolmogorov
velocity.  We consider moderately turbulent flows, with Taylor-scale
Reynolds number $\mathrm{Re}_\lambda\!=\!\sqrt{15}u_\mathrm{rms}/(\nu
\epsilon)^{1/2} \approx 70\!-\!180$.

For non-spherical particles we find a remarkable alignment of the
swimming direction with the local velocity field quantified by the
statistics of the angle, $\theta_u$, between $\bm n$ and $\bm u$.
Figure~\ref{fig1}a shows that $\langle \cos\theta_u\rangle\neq 0$
provided the particle is not spherical ($\Lambda\neq 0$) and active
($\Phi>0$) ($\langle[\ldots]\rangle$ denotes the average over particle
positions).  Data suggest that $\langle \cos \theta_u \rangle \propto
\Lambda \Phi$ (inset of Fig.~\ref{fig1}a) at least for small $\Phi$
and $\Lambda$, with some deviations from linear behavior for
$|\Lambda|\to 1$.  We remark that such alignment depends on the
particle shape: it is ``polar-like'' for elongated particles
($\Lambda>0$) and anti-polar for disk-like ones ($\Lambda< 0$).  Thus,
on average, rod-like particles swim along the underlying flow velocity
while disk-like ones against it.

The variance of $\cos\theta_u$ (Fig.~\ref{fig1}b) displays a
non-trivial dependence on $\Lambda$ while it is almost insensitive to
$\Phi$, except around $\Lambda \to 1$, where it slightly decreases
with $\Phi$. The different behavior for disk-(rod-)like particles
reflects qualitative differences in the probability density function
(PDF) of $\cos\theta_u$.  For disks (bottom inset), the PDF of
$\cos\theta_u$ displays a peak that gradually moves from 0 (swimming
normal to velocity) to negative values at increasing
$\Phi$. Conversely, for rods (top inset) the PDF is bimodal at $\pm 1$
for $\Phi=0$ with a progressive bias in favor of $+1$ peak at
increasing $\Phi$.  Thus elongated particles ($\Lambda>0$) align with
the local fluid velocity for any $\Phi$ but the alignment changes from
nematic to polar upon increasing $\Phi$.

\begin{figure}[t!]
\centering
\includegraphics[width=0.9\columnwidth]{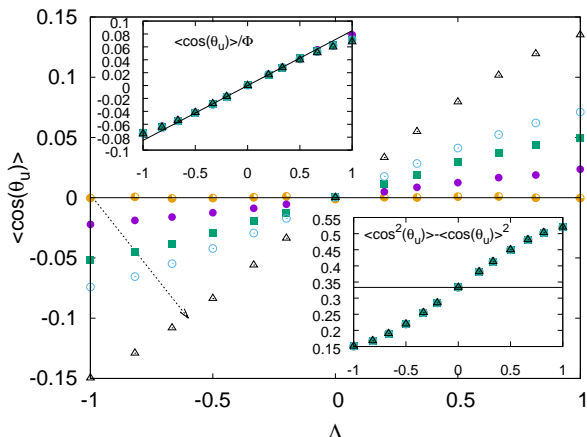}
\vspace{-0.5truecm}
\caption{(Color online) Statistics of orientation obtained by
  simulations of the stochastic model for $\mathrm{Ku}=10$. Main
  panel: $\langle \cos\theta_u\rangle$ vs $\Lambda$ for different
  swimming speeds, $\Phi_s=0,0.3,0.7,1,2$ as indicated by the arrow,
  Top inset: $\langle \cos\theta_u\rangle/\Phi_s$ vs $\Lambda$. Bottom
  inset: variance of $\cos\theta_u$.
  \label{fig2}}
\end{figure}

To rationalize the above observations, we now consider a statistical
model for the velocity field, ${\bm u}({\bm x},t)$, which allows for
analytical treatments.  As detailed in \cite{gustavsson2016review}
(see Sect. I.A in the Supplemental Material (SM)\footnote{See {S}upplemental {M}aterial [url], which includes Refs.\cite{Fal01,Gustavsson2019,Fri97,Cal09,Gus11a},  for a full description of the statistical model, for details on the perturbative calculations, and for results on the alignment with vorticity.}), we
consider a (single scale, single time) random Gaussian velocity field
parameterized by typical flow speed $u_f$ with correlation length,
$\ell_f$, and time, $\tau_f$.  We introduce the additional
dimensionless number, namely the Kubo number $\mathrm{Ku} = u_f
\tau_f/\ell_f$, quantifying how rapidly the fluid velocity
fluctuates. Figure~\ref{fig2} displays the statistics of alignment
obtained from a numerical simulation of the stochastic model. The
agreement with the results obtained in turbulent flows
(Fig.~\ref{fig1}) is remarkable, demonstrating that the alignment is a
robust \textit{kinematical} phenomenon, i.e. independent of the
dynamics producing the flow.  This is in contrast with the known
alignment observed for elongated swimmers with the local vorticity
$\bm \omega(\bm x,t)=\bm \nabla \times \bm u(\bm x,t)$
\cite{zhan2014accumulation,pujara2018rotations}, which is absent in
the stochastic flow. Indeed, the origin of alignment with vorticity is
\textit{dynamical} as discussed in \cite{pumir2011orientation} for
elongated tracers and stems from the formal similarity of
Eq.~(\ref{eq:dotn}) for $\Lambda=1$ with the Lagrangian dynamics of
vorticity. See Sect. IV in SM \cite{Note1} for further considerations.

\begin{figure}[b!]
\centering
\includegraphics[width=.9\columnwidth]{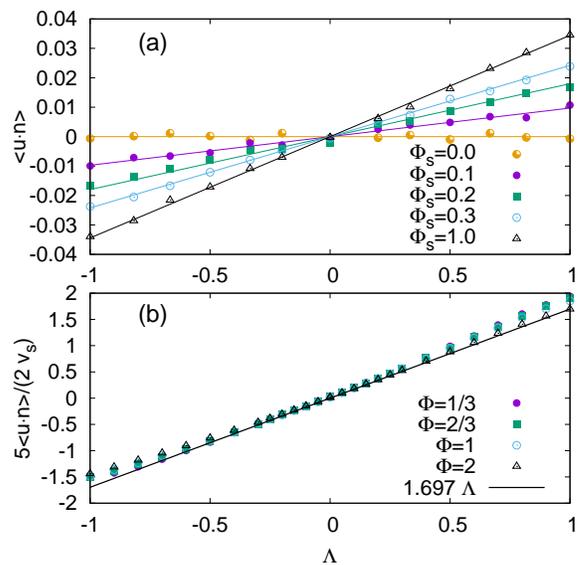}
\vspace{-0.5truecm}
\caption{(Color online) Comparison between theoretical predictions and
  simulations for both the statistical model and DNS. (a) $\langle \bm
  n\cdot \bm u\rangle$ vs $\Lambda$ for different $\Phi_s$ obtained
  numerically for the statistical model with $\mathrm{Ku}=0.1$
  (symbols) compared with the theoretical prediction
  (\ref{eq:alignment_final}) (solid lines).  (b) $5\langle(\bm n \cdot
  \bm u)\rangle /(2v_s)$ vs $\Lambda$ for different $\Phi=v_s/u_\eta$
  for the DNS at $Re_\lambda\approx 68$. The solid line represents the
  prediction (\ref{eq:6}), $1.697\Lambda$ with the numerical prefactor
  was numerically obtained evaluating the strain correlation function
  along tracer trajectories.
\label{fig3}}
\end{figure}

The advantage of the stochastic model is that it allows for reaching
an analytical understanding of the basic mechanism for the
alignment. In particular, we study the statistics of $(\bm n\cdot\bm
u)$ instead of $(\bm n\cdot\bm u)/|\bm u|\!=\! \cos\theta_u$, as they
convey the same qualitative information on alignment
(Fig.~\ref{fig3}), and are easier to handle.  The main difficulty in
analyzing Eqs.~(\ref{eq:dotx}-\ref{eq:dotn}) lies in their non-linear
dependence on the particle position. Such a hindrance can be overcome
in perturbation theory, by iteratively improving approximations for
the particle trajectory, a technique successfully employed to analyze
inertial particles \cite{gustavsson2016review} and gyrotactic swimmers
\cite{gustavsson2016gyrotaxis}.  This corresponds to an expansion in
the Kubo number~\cite{gustavsson2016review}.

In the following we briefly outline the main steps, detailed
calculations can be found in Sect. II.A of SM \cite{Note1}.  To apply
perturbation theory we introduce dimensionless variables with
$t=t'\tau_f$, $x = x'\ell_f$, and $u = u' u_f$, in terms of which
Eqs.~(\ref{eq:dotx}-\ref{eq:dotn}) read: $\dot{\bm x}'=\mathrm{Ku}\bm
u'+\Phi_s\bm n$ and $\dot{\bm n}=\mathrm{Ku}\bm J'(\bm n)$, with
swimming number $\Phi_s=v_s\tau_f/\ell_f$. The above equations imply
that, for $\mathrm{Ku}\!=\!0$, the particle paths are simply ${{\bm
    x}'}^{({\rm d})}_{t'} =\bm x_0'+\Phi_s \bm n_0 t'$, where ${(d)}$
denotes the zeroth order (deterministic) solution.  We can now write
$\bm x_{t'}'\!  =\! {{\bm x}'}^{({\rm d})}_{t'} + \delta \bm x_{t'}$
and expand Eqs.~(\ref{eq:dotx}-\ref{eq:dotn}) to the desired order in
$\delta \bm x_{t'}$, leading to an expansion in $\mathrm{Ku}$ at fixed
$\Phi_s$ \cite{gustavsson2016review,Note1}. The result is then averaged
using the known correlation functions of the (Gaussian) velocity field
and its derivatives. Using flow isotropy, homogeneity, and
incompressibility, the stationary-state average of the scalar product
between $\bm n$ and $\bm u$ takes the (dimensional) form:
\begin{eqnarray}
\langle \bm n\cdot\bm u \rangle\!=\!-{d\Lambda}\!\! \int_0^t\!\!\!\!{\rm d}t_1
\partial_R C_\parallel(R, t_1)\big|_{\bm R = \bm x_{t_1}^{(d)}}\,,
\label{eq:alignment_final}
\end{eqnarray}
$d$ being the spatial dimension and $R\!=\!|\bm R|$. For the
stochastic flow, the longitudinal velocity covariance takes the form
$C_\parallel(R,t) \!\equiv \!\langle(\bm u(\bm x+\bm R,t)\cdot\hat{\bm
  R})(\bm u(\bm x,0)\cdot\bm{\hat R})\rangle
\!=\!\exp[{-(R^2/{2\ell_f^2} + |t|/\tau_f)}]/d$.  Substituting it in
Eq.~(\ref{eq:alignment_final}) and using ${\bm x}_{t_1}^{(\rm
  d)}\!=\!\Phi_s\bm n_0 t_1$, the integral can be easily computed (see
Sect. II.C in SM \cite{Note1}), yielding for $\Phi_s\!\ll\! 1$
\begin{equation}
\langle \bm n\cdot\bm u\rangle \simeq u_f \Lambda \mathrm{Ku} \Phi_s\,,
\label{eq:prediction}
\end{equation}
which agrees well with the numerically obtained scaling of
$\cos\theta_u$ in terms of $\Lambda$ and $\Phi$ (inset of
Fig.~\ref{fig1}a and Fig.~\ref{fig2}).  Figure~\ref{fig3}a shows that,
for $\mathrm{Ku}\ll 1$, statistical-model simulations perfectly agrees
with the theoretical prediction.

Neglecting vorticity in (\ref{eq:dotn}), in the limit of small swimming speeds
and $|\Lambda|$, an expansion similar to that used in \cite{Gus16b} can be performed yielding (see Sect. V in SM \cite{Note1})
\begin{equation}
\langle \bm n \cdot \bm u\rangle \!=\! \frac{2\Lambda v_s}{d+2}\int_0^t \!\!dt_1 t_1 \Tr\langle \mathbb{S}(\bm x^L_{t_1},t_1)\mathbb{S}(\bm x^L_{0},0)\rangle,
\label{eq:6}
\end{equation}
which expresses $\langle \bm n\cdot\bm u\rangle$ in terms of the
correlation function of the strain along Lagrangian trajectories, ${\bm x}^L_t$, i.e. corresponding to the dynamics
(\ref{eq:dotx}) with $v_s=0$. Note that the above expression, being free from any assumption on the flow statistics, only requires $v_s$ and $\Lambda$ to be small  and  should therefore be valid
for generic flows and  Kubo numbers (see Sect. V in SM
\cite{Note1}).  We measured $\Tr\langle \mathbb{S}(\bm
x^L_{t},t)\mathbb{S}(\bm x^L_{0},0)\rangle$ along Lagrangian trajectories in DNS
and numerically computed the integral in (\ref{eq:6}) obtaining a
prediction for $\langle \bm n\cdot\bm u\rangle$ that agrees well with
the numerical data, at least for not too large $\Phi$ and $\Lambda$
(Fig.~\ref{fig3}b).

The physical meaning of Eq.~(\ref{eq:alignment_final}) is as follows:
The alignment results from the non-zero correlation between velocity
and its gradients at different times (this is $\partial_R
C_\parallel(R,t\!-\!t_1)$). However, such correlation brings a
non-zero contribution only if swimming ($v_s\neq 0$) and
non-sphericity ($\Lambda\neq 0$) are present. Essentially swimming in
the instantaneous direction breaks the fore-aft symmetry and the
dynamics of $\bm n$ is no longer identical to the dynamics of $-\bm
n$, leading to a non-zero value for $\langle \bm n \cdot \bm u\rangle$
(see Sect. II.C in SM \cite{Note1} for further considerations).  For
the second moment, the situation is different, to order
$\mathrm{Ku}^2$ and for $\Phi \ll 1$ we find (see Sect. III in SM
\cite{Note1})
\begin{equation}
\langle(\bm n\cdot\bm u)^2\rangle/u_f^2\!\sim\!\frac{1}{d}\!+\!\frac{\mathrm{Ku}^2\Lambda}{d}\!+\!\frac{\mathrm{Ku}^2\Phi^2\Lambda}{2d} (2d\Lambda\!-\!11)\,,
\end{equation}
which depends on $\Lambda$ also for non-swimming particles, as
confirmed by simulations.

In the statistical model, there is a single time scale for both
velocity and its gradients. Conversely, in turbulence there is a time
scale separation between them controlled by
$\mathrm{Re}_\lambda\!\approx\! T/\tau_\eta$ ($T$ being the integral
timescale). Thus, for strong turbulence, the dynamics of the
orientation, ruled by velocity gradients, will vary over time scales
($\sim\! \tau_\eta$) much faster than the correlation time of the
velocity ($\sim\!T$), possibly depleting the alignment.  This is
confirmed in the inset of Fig.~\ref{fig1}c showing that $\langle
\cos\theta_u\rangle \sim \mathrm{Re}_\lambda^{-1/2}$. This
$Re_\lambda^{-1/2}$ scaling can be rationalized as follows. The
statistical model calculations predict that, for given $\mathrm{Ku}$,
alignment only depends on $\Lambda$ and $\Phi_s$.  Thus we need to map
the swimming parameter of the model on that used in turbulence.
Following Ref.~\cite{gustavsson2016gyrotaxis} (see also Sect. I.A in
SM \cite{Note1}) the statistical model length ($\ell_f$) and time
($\tau_f$) scales should be related to the Taylor length scale,
$\lambda\propto u_{\mathrm{rms}}\tau_\eta$ and $\tau_\eta$,
respectively, being the scales relevant to the gradients. Therefore,
the swimming number to be used to compare DNS with the statistical
model should be based on the r.m.s. velocity, indeed $\Phi_s=
v_s\tau_\eta/\lambda\propto v_s/u_{\mathrm{rms}}$ (see also
\cite{borgnino2018shape} for related considerations), while we used
$\Phi=v_s/u_\eta$. The two swimming numbers are thus related by
$\Phi_s\propto \Phi (u_\eta/u_{\mathrm{rms}})\propto Re_\lambda^{-1/2}
\Phi$, which explains the scaling observed in Fig.~\ref{fig1}c.  DNS
results (not shown) confirm that for fixed $\Phi_s$ the alignment
statistics is independent of $Re_\lambda$.  Thus, alignment can be
important also for high $Re_\lambda$ flows provided the particle speed
is a fraction of the large scale velocity. Such large speeds can be
attained by swimmers larger than the Kolmogorov scale, for which
Eqs.~(\ref{eq:dotx}-\ref{eq:dotn}) may still be valid, provided the
Stokes number defined on the particle scale is small enough, as
recently found in finite-size fibers \cite{loshe}.

In general, alignment is expected to be important whenever turbulence
is moderate, i.e. in velocity fields with not too separated scales of
motion, as commonly found in environmental, laboratory and biomedical
fluids. In marine environments with calm water the Kolmogorov velocity
is in the order of $u_\eta\!\approx\!  300\!-\!1000 \mu m/s$
\cite{kiorboe2008} while bacterial speeds range in $v_s\!\approx\!
30\!-\!300 \mu m/s$ \cite{barbara2003marine} consequently $\Phi\approx
0.05- 1$. Hence, depending on the Reynolds number, alignment can be
substantial.  Alignment could be relevant to models for light
scattering in aquatic environments \cite{boss,seymour2011microbizal}
especially considering that most of motile microorganisms are
elongated \cite{boss}, and for the encounter rates of aquatic
microorganisms \cite{kiorboe2008,visser}. Further, analogously to the
findings in steady shear flows \cite{rusconi2014}, flow reorientation
may alter the chemotactic efficiency.  For instance, flow-induced
alignment could be particularly relevant to marine bacteria, many of
which perform a run-reverse cycle in which the orientation is
unchanged while the swimming velocity is reversed
\cite{stocker2012ecology}.

Preliminary studies, to be discussed elsewhere, show that alignment
persists also in the presence of a non-homogeneous mean flow. In this
case, alignment may dramatically impact the dispersal properties along
and transverse to the mean flow similarly to what is observed in
steady porous flows \cite{dehkharghani2019bacterial}, with
implications for groundwater filtration and remediation, and
biomedical fluids.  Remarkably, the experiments in
\cite{dehkharghani2019bacterial} demonstrated preferential alignment
of the bacterial swimming direction with the local flow in analogy
with our findings.  It would be then interesting to study alignment in
the limit of steady flows, i.e. in the $\mathrm{Ku}\to \infty$ limit,
to understand whether the physical mechanism for alignment is the same
of that we found in unsteady flows. This is however beyond the scope
of the present Letter.

Finally, we observe that nontrivial correlations between flow velocity
and individual bacterial orientation have been reported in dense
suspensions \cite{sokolov2007,ryan2016swarming}, where the
self-generated flow is in the order of $\sim 50-100 \mu m/s$ with
correlation length of $30-100\mu m$, while bacteria swim at speed
$\sim 15-20\mu m/s$ with a size of $\sim 2\mu m$
\cite{dombrowski2004self,sokolov2007}. With swimming numbers in the
order of $\approx 0.15-0.4$, it is tempting to speculate that the
alignment here discussed could be an important effect. However, this
needs to be tested because steric and hydrodynamic interactions, here
neglected, play a major role.

Summarizing, we found that (disk-)rod-like active particles swimming in a
moderately turbulent background flow tend to preferentially align
their swimming direction (anti) parallel  to the underlying flow velocity.
We showed that such an alignment has a kinematical
origin and analytically found its roots in the time
correlations between velocity and its gradients along particle paths
together with the fore-aft symmetry breaking induced by
 swimming.  Our study expands on the possible
non-trivial behaviors of microswimmers in an external flow
\cite{guasto2012,rusconi2015} from the simple cases of pipe or shear
flows \cite{torney2007transport,zottl2012} to
  realistic unsteady turbulent and chaotic flows.

\begin{acknowledgments}
We acknowledge useful discussions with R.  Stocker.  MB, GB and FDL
acknowledge support by the {\it Departments of Excellence} grant
(MIUR).  BM and KG acknowledge Knut and Alice Wallenberg Foundation,
grant no.~KAW~2014.0048, and Vetenskapsr\aa{}det, grant no.~2017-3865.
CINECA is acknowledged for computing resources, within the INFN-Cineca
agreement INF18-fldturb and the Iscra-C GyATuS grant.
\end{acknowledgments}

%

\end{document}